# Mass-velocity and size-velocity distributions of ejecta cloud from shock-loaded tin surface using atomistic simulations

O. Durand, L. Soulard

CEA, DAM, DIF, F-91297 Arpajon, France

**Abstract**: The mass (volume and areal densities) versus velocity as well as the size versus velocity distributions of a shock-induced cloud of particles are investigated using large scale molecular dynamics (MD) simulations. A generic three-dimensional tin crystal with a sinusoidal free surface roughness (single wavelength) is set in contact with vacuum and shock-loaded so that it melts directly on shock. At the reflection of the shock wave onto the perturbations of the free surface, two-dimensional sheets/jets of liquid metal are ejected. The simulations show that the distributions may be described by an analytical model based on the propagation of a fragmentation zone, from the tip of the sheets to the free surface, within which the kinetic energy of the atoms decreases as this zone comes closer to the free surface on late times. As this kinetic energy drives (i) the (self-similar) expansion of the zone once it has broken away from the sheet and (ii) the average size of the particles which result from fragmentation in the zone, the ejected mass and the average size of the particles progressively increase in the cloud as fragmentation occurs closer to the free surface. Though relative to nanometric scales, our model may help in the analysis of experimental profiles.

# I. INTRODUCTION :

The particulate (ejecta) emission that occurs in front of the free surface of a shock-loaded material is a phenomenon that has paid ongoing attention for several decades in the shock compression community. Indeed, it can not only perturb the diagnostics (electrical and optical) implemented to characterize the dynamics of the material (for instance the measurement of its free surface velocity) but it can also be a source of inhibition in applications like inertial confinement fusion [1]. The ultimate goal is to implement an ejecta source model that is able to predict both the total amount of ejected mass and the distribution of this mass, i.e. the distributions of the ejecta in size and velocity. Past theoretical and experimental efforts have shown that surface imperfections of machined material (in particular two-dimensional periodic features for metals) are the main cause of ejecta production [2-6].

Experimentally and numerically, the ejecta cloud has been and continues to be characterized mainly globally, at the continuum level, through the measure of the total amount of ejected mass as a function of the velocity [7-16] and by using the conservation laws of hydrodynamics in computations [1, 17-18]. From these studies, is now well admitted that the phenomenon is a limiting case of Richtmyer-Meshkov instabilities (RMIs) [1, 19-20].

Investigating experimentally the size distribution of the particles inside the cloud is hard to achieve. Relatively few data are available [21-23], and ongoing and long-term developments of specific tools [14, 24] have to be undertaken. Theoretically and numerically, if it seems possible to estimate a distribution of characteristic particle sizes for small sinusoidal perturbations [16, 20], models and hydrocodes fail for the moment to provide correlation between size and velocity of the particles inside the cloud. In particular the classical hydrodynamics approaches suffer a lack of modeling and characterization of the rheological behavior of metals after shock loading; and data on surface tension and viscosity are needed. To our knowledge, only recent MD simulations performed on atomistic systems allowed to provide directly from computations ejecta size distributions [25-27]. Particulate computations are indeed at the good scale to provide information on the microscopic mechanisms of plastic deformation and/or fragmentation of metals under melting or release melting conditions [25-31]. However, the main objection on these simulations is that the scales and times involved are smaller than those in real structures by at least 3 or 4 orders of magnitude; and in the absence of reliable data, it is effectively not yet possible to confirm the results with experiments or models.

In order to prove the interest of such atomistic approach to deal with ejection phenomenon, there is a need to compare the results with existing data; i.e. the continuum aspects need to be studied. This is the first goal of this study. The second one is to show that MD simulations may be very helpful for modeling the RMIs fragmentation since they allow to identify physical parameters which are also calculable at macroscopic scale.

## II. CONDITIONS OF COMPUTATION

We briefly review in this section our conditions of computation; details can be found in Ref. [25-26]. The simulations are performed with STAMP code and TERA-100 computer of CEA, using typically between 2000 and 4000 cores. A generic configuration consisting in a three-dimensional (3D) crystal with a sinusoidal free surface roughness (3 arches) is set in contact with vacuum and shock-loaded so that it melts directly on shock. The time step is 1 fs and periodic conditions are applied in the $O_y$ and $O_z$ directions while the shock propagates in the $O_x$ direction (free condition). The initial material is a tin crystal in BCC phase with the [100] direction parallel to the $O_x$ axis ($L_x$ axis). The lattice constant $a_0$ is 3.700 Å and we use the EAM potential developed by Sapozhnikov et al. [32]. The crystal is made up of 300 × 400 × 250 elementary cells $a_0$ (dimensions $L_x \times L_y \times L_z$: 111 nm × 148 nm × 92.5 nm) and contains about $60 \times 10^6$ atoms. The thermodynamic properties of the shock-loaded crystal (compression ratio of the density, particle velocity, shock pressure and temperature) are reported in Table I. The compression ratio ensures that the crystal is above its melting point, in a purely liquid phase.

The wavelength $\lambda$ of the roughness is kept constant and equal to 30.8 nm ($L_z$/3) and 3 values of amplitude $h$ are tested: 2, 5, and 10 nm, as indicated in Fig. 1. The corresponding wave number/amplitude product $kh$ ($k = 2\pi/\lambda$) for each amplitude is: ~ 0.42, 1.05, and 2.1 for $h$=2, 5, and 10 nm, respectively. From a continuum point of view, both linear ($kh < 1$) and nonlinear ($kh > 1$) regimes of RMIs are covered.

In order to compare our simulations with existing simulations/models and experiments, we use the same methodology as the one developed by Dimonte et al. [1]. We define the distance in the cloud relative to the position of an unperturbed (flat) interface defined as the free surface. The position and velocity of the free surface are calculated in a separate simulation on a tin crystal with no roughness ($h$=0). The crystal is made up of 300 × 40 × 40 elementary cells $a_0$ and the position of the free surface of this crystal at the beginning of the calculations is also indicated in Fig. 1 in dashed line. With our calculation conditions, the free surface velocity is measured to be 3320 m/s, nearly twice the particle velocity (1570 m/s) as expected, if isentrope and Hugoniot curves are symmetrical.

The simulations are performed using the hybrid method we previously developed and which is described in Ref. [25]. The analysis tools implemented for the characterization of the particles/aggregates/ejecta (these terms refer to the same thing) in size, velocity, position and mass are also described in Ref. [25-26]. In particular, the system is divided in voxels and the Hoshen-Kopelman algorithm [33] is used to label the atoms that belong to the same aggregate. A binning technique is used in order to improve the statistics of measurement of the particle size distributions [34].

## III. THE CONTINUUM POINT OF VIEW

As mentioned earlier, we aim at studying in this section the continuum aspects of ejection with MD simulations. A direct comparison with existing experiments is not possible. Indeed, our Sn crystal is driven to a very high supported shock wave pressure (P > 60 GPa) to reach directly Sn liquefaction (and avoid in particular to deal with melting in release kinetics); and experimentally, such high pressure cannot be reached either by powder gun (supported shock-waves) or high explosive (unsupported shock-waves) experiments [14]. Typically, pressures of ~ 27 GPa were obtained with explosively driven experiments in references [7-14]. Therefore, what seems the most informative for us is to compare our results with some recently developed ejecta models [1, 19], based on the nonlinear evolution of RMI [1, 19]. Although these models have not been tested at extreme pressures, it is interesting to see how our results compare with their conclusions (and since no data exist at extreme regimes, our results may even form a first data basis for the validation of ejecta models at high pressure). We use in particular the model developed in the study of Buttler et al. [19]. In this study, the asymptotic limit $V_{limit}$ of bubble velocity as a function of time $t$ derives from the Mikaelian's model [35]: $V_{limit} = 2/(3kt)$ (on late times, it is independent of the initial perturbation amplitudes); and the maximum velocity attainable by the spikes is $V_{Spike}^{max} = \sqrt{3} V_{Spike}^{0}$, with:

$$V_{Spike}^{0} = \left(1 - \frac{V_{FS}}{2V_{Shock}}\right)\left(\frac{1}{1+\left(\frac{\eta_0 k}{2}\right)^2}\right)\eta_0 k V_{FS}, \quad (1)$$

$V_{Shock}$ being the shock velocity ~ 4986 m/s (with $V_{FS}$ ~ 3320 m/s) and $\eta_0$ the perturbation amplitude (here $\eta_0 = h/2$). This expression includes portions of Mikaelian's model [35] with linear [36] and nonlinear [19] corrections.

In Fig. 2a, we show side views of the ejected sheets of mass for the 3 roughness amplitudes at a given time (200 ps) and in Fig. 2b we show the corresponding distributions of volume density versus distance. As mentioned earlier, this distance is defined relative to the position of the (flat) free surface measured on a separate simulation at the same time. Adopting the same notation as in Ref. [1] in RMIs description, we define the region of void between the bottom of the jet (once the phase inversion of the defect has occurred) and the free surface (which position is represented by a white dashed line in Fig.2a) as a bubble, and the region of the jet beyond the free surface as a spike. The distance between the bottom of the jet and the free surface is defined as the bubble amplitude $h_{Bu}$ and the distance between the tip of the spike (or jet or sheet, these terms refer to the same thing) and the free surface is defined as the spike amplitude $h_{Sp}$. These distances are quantitatively determined by considering in Fig. 2b the bottom of the jet there where the volume density is 99% of the maximum volume density (bulk density) and the tip of the jet there where the volume density is at its secondary maximum (peak). This peak corresponds to a local concentration of matter caused by the effects of surface tension which minimize the energy of the jet at its tip, as already mentioned in Ref. [26]. Note that the bulk density ~ 6500 mg/cm$^3$ is slightly below the nominal density of Sn because the metal releases and its temperature is high. One important thing to note is that the volume density profile of Fig. 2b does not give any information on the state of the ejected matter. The latter may be in a continuum (described by fluid mechanics) or in an ejecta particle state. It just gives information on the amount of ejected matter per unit volume.

In Fig. 3 and 4, we report the temporal evolution of the bubble and spike amplitudes (Fig. 3a and 4a respectively) and the temporal evolution of their derivative, the bubble and spike velocities (Fig. 3b and 4b respectively), for the 3 roughness amplitudes. In Fig. 3b, we report also the asymptotic limit $V_{limit}$ of bubble velocity from Mikaelian's model [35]; and we compare it in the insert to the bubble velocities on late times. We note that while at the beginning of the process (up to 100 ps) the bubble grows negatively very strongly, then the velocity decays following the expected $1/t$ law but on long times it goes to an asymptotic value (relative to the free surface velocity) which is finite and constant, of the order of ~ - 12 m/s, -31 m/s and -40 m/s for $h$ = 2, 5 and 10 nm respectively. This implies that the bubble amplitude does not saturate. We explain this difference with the expected limit of zero in the model ($V_{limit} \to 0$ as t $\to \infty$) first by the finite dimensions of our system. Indeed, when the release waves issued from the reflection of the shockwave onto the free surface reach the backside of our crystal, the system comes off the piston that was applied to support the shock. Then the crystal becomes isolated, and when the sheets are ejected, it loses a very small part of its velocity (momentum conservation) that can no more be compensated by the piston, unlike what occurs in hydro computations [1, 17-18] and experiments [2-16] where the samples have "infinite" (very large) dimensions. Other sources may also explain this difference; in particular the conditions of simulation (supported shockwave at extreme pressure).

While the bubbles grow negatively, the spikes grow in the opposite direction, resulting in an elongation of the jet (see Fig. 4). Conversely to the bubbles, we observe that the spike velocity (i.e. the jet tip velocity) is constant from the very beginning of the process. The spike velocities (with respect to the free surface) are ~ 311 m/s, 1134 m/s and 1920 m/s, and they correspond to spike velocities in the laboratory frame ($V_{Sp} + V_{FS}$) of ~ 3631 m/s, 4454 m/s and 5240 m/s for $h$ = 2, 5 and 10 nm respectively. From eq. (1), the maximum velocities attainable with the model for $h$ = 2, 5 and 10 nm, are ~ 795 m/s, 1880 m/s and 3153 m/s respectively; therefore the model always over-predicts our velocities by about 40% to 60%. Applied to the Sn experiments at lower pressures (~ 27 GPa) [19], the model tends also to over-predict the spike velocities for $kh$ product ≥ 1. However the difference is at maximum by about 20%, and the model predicts the spike velocity within 2% for $kh$ product ≤ 0.5. In our case, the discrepancy may be attributed to the fact that our simulations take inherently into account the effects of viscosity and surface tension (through the potential), which is not the case for the current RMIs studies. As for the bubbles, our specific conditions of simulation may also be another source of difference.

The large difference of velocities between bubble and spike makes the RMIs profile highly asymmetric (relative to the position of the free surface). This is displayed in Fig. 5 where we report the temporal evolution of the bubble and spike velocities in the laboratory frame ($V_{Bu/Sp} + V_{FS}$) normalized to the free surface velocity. Note that the negative character of $V_{Bu}$ appears more clearly by writing $V_{FS} + V_{Bu} = V_{FS} - |V_{Bu}|$. Whatever the RMIs regime, the normalized bubble velocities are close to 1 (between 0.985 and 0.995) while the normalized spike velocities are of the order of 1.1, 1.35 and go up to 1.57 for $h$ = 2, 5 and 10 nm respectively. These values are of the same order of magnitude as experimental ratios obtained in literature [7-14] with similar $kh$ products, although our conditions of simulations are different.

In Fig. 6a, we represent, for a range of times from 100 ps to 1300 ps, the velocity distribution inside the ejected sheets (in the laboratory frame) as a function of the distance (relative to the free surface), for $h$ = 5 nm. The same generic behavior was observed for the three amplitudes. In this figure, we also indicate the free surface velocity $V_{FS}$ as well as the spike velocity in the laboratory frame ($V_{FS} + V_{Sp}$). The latter is constant whatever the time, as already shown in Fig. 4b. For a given time, we observe that the velocity inside the sheet increases linearly with the distance from $V_{FS}$ to $V_{FS} + V_{Sp}$; consequently, the expansion rate (or gradient velocity) of the sheet is constant all over its length and is equal to $\dot{\varepsilon} = V_{Sp}/h_{Sp}$. Since $V_{Sp}$ is constant and $h_{Sp}$ increases linearly with time (see Fig. 4a), $\varepsilon$ decreases as $1/t$. This is observable in Fig. 6b where we represent the temporal evolution of the expansion rate for the three amplitudes. Note that we deal here with expansion rates (> $10^9$ s$^{-1}$) which are much higher than usual experimental values (~ $10^5$-$10^6$ s$^{-1}$), because we apply typical velocity differences of about 1000 m/s between ends separated of only several hundreds of nm (against several millimeters in experiments and hydro simulations).

In Fig. 7, we report for a range of times from 200 ps (beginning of the ejection/fragmentation process) to 1500 ps (end of the process), the profiles of the ejected volume density $\rho_{ejecta}$ (Fig. 7a) and of the cumulative areal density $\sigma_{ejecta}$ (Fig. 7b and 7c, in linear and semi-logarithmic scales respectively), as a function of the distance from the free surface, for $h$ = 2 nm (the same generic behavior was observed for the three amplitudes). The areal density profile is obtained by integrating the volume density profile along the jet, starting at the tip. At a given longitudinal position $x < h_{Sp}$ and time $t$, it is given by:

$$\sigma_{ejecta}(x,t) = \int_{h_{Sp}(t)}^{x} \rho_{ejecta}(x',t)dx'. \qquad (2)$$

We note that the volume profile changes its morphology during the process. It is smooth and continuous in the early times from the bottom of the jet to the tip; and on late times (> 1 ns) it becomes steady and exhibits a much stiffer edge which allows to clearly distinguish the bulk from the ejected mass. We note also in the insert of Fig. 7a that beyond typically 1 ns, the amount of volume density measured at the bottom of the bubble, i.e. at the bulk edge, is of the order of 70 to 100 mg/cm$^3$. This amount is of the same order of magnitude as experimental values [7-14] (although our conditions of simulations are different from experiments). Consequently, the cumulative areal density profile exhibits also a clear change of morphology at the end of the process (Fig. 7b and 7c). Conversely to the volume density, the amounts of areal densities cannot be compared to experiments since we integrate our volume density profile on hundreds of nm against several millimeters in experiments.

The total (maximum) amount of areal density ejected ($\sigma_{ejecta}^{max}$) is evaluated by integrating $\rho_{ejecta}$ from the tip to the bottom of the jet, i.e. by setting $x$= -|$h_{Bu}$(t)| in eq. (2) [1]. Since the profile of $\rho_{ejecta}$ is steady on late times, the amount of areal density ejected does not vary anymore on late times too. In Fig. 8a, we report the temporal evolution of $\sigma_{ejecta}^{max}$ for the three amplitudes and, in Fig. 8b, we report the error of the measure on the value obtained at the last time (1500 ps, 1300 ps and 1100 ps for h= 2, 5 and 10 nm respectively). We note that an error of ~ 90 % for $h$ = 2 and 5 nm and up to ~ 160% for $h$ = 2 nm can be made, depending on the time (here around 500-600 ps) at which the measurement is performed. The temporal aspects in the determination of the amount of ejected areal density are therefore very important. This definition of a fixed and no more varying value of areal density with time may seem in contradiction with a boundary of integration -|$h_{Bu}$(t)| which varies with time (it grows at constant velocity, see Fig. 3a). We will see in the next section that the process of fragmentation of the jet makes this definition still relevant.

Thus, in this section we have shown that, although our conditions of simulation are rather extreme (supported shockwave at extreme pressure) and dissimilar from experiments and model, the velocities and amount of volume density ejected still remain of the same order of magnitude. Therefore, we are confident in the capability of our MD simulations to reproduce the continuum aspects of ejection, in particular within the framework of RMI. A more detailed study between model and MD simulations should be performed, but it is out of scope of the present study.

Furthermore, MD simulations offer another possibility: they allow to identify all the parameters that contribute to the fragmentation of the jets. We are going now to study in detail this mechanism and see how it can explain the bubble and spike behavior as well as the profiles of volume and areal densities.

## IV. FRAGMENTATION OF THE SHEETS AND MODEL

### A. FRAGMENTATION OF THE SHEETS

As already studied in Ref. [25-26], when the ejected sheets get thin down to a critical value $e_c$ of about 4 nm, they become highly unstable; and their thickness undergoes stochastic and very local fluctuations from which the nucleation of some first holes may start [37]. This is observable in Fig. 9 where we present top and side views of the upper ejected sheet (representative of the 3 sheets) for the 3 amplitudes, at 150 ps, 90 ps, and 80 ps for $h$ = 2, 5 and 10 nm respectively. Note that the thinning down of the sheets occurs in a region of width $L_F$ in the ($O_x$) direction of elongation of the sheets (starting at the tip) which is as larger as the initial roughness amplitude is increased, due to the initial expansion rate. The characteristic time $\tau$ taken by a hole to remain open and become a pore of void which will later grow, coalesce with other pores and lead to the fragmentation of the sheet [26] is driven by the effects of surface tension. By dimensional analysis [38]: $\tau \propto \sqrt{\dfrac{\rho R^3}{\sigma}}$, where $R = e_c/2$, $\rho$ is the volume density of Sn within the jet (~ 6500 kg/m$^3$ like in the bulk) and $\sigma$ is the surface tension of Sn (~ 0.57 N/m); in our case, $\tau$ ~ 10-20 ps. This timescale is confirmed by comparing the top views of Fig. 9 with the top views (for the same systems and computation conditions) of Fig. 3 in Ref. [26] taken ~ 20-40 ps later. Some pores of void are then clearly distinguishable.

Let us now analyze the evolution of the sheets as a function of time. In the following, we concentrate our study on the linear regime of RMIs, with $h$= 2 nm, knowing that the observations are the same whatever the regime. At different times from 200 ps to 700 ps, we show in Fig. 10 oblique views of the system and, in Fig. 11, we show top and side views of the upper sheet. We observe that the thinning down of the sheets (at ~ 4 nm) and, as a consequence, the appearance of holes, occurs closer and closer to the free surface. By considering the edge of the holes the closest to the free surface, it is possible to define (roughly) a front of fragmentation, as indicated by a white continuous line in the snapshots of Fig. 11. This front counter-propagates from a given point in the sheet in the early times (for the case $h$ = 2 nm, the starting point is located near the tip at 200 ps) towards the free surface (and the bulk). We found that the velocity of propagation of the front is constant whatever the regime; for $h$ = 2 nm, it is about -80 m/s. Behind this front, the sheet breaks up.

The propagation of the front is also detectable in Fig. 12a where we report for the same times as above the velocity profiles inside the sheet as a function of the distance (from the free surface). As for the case $h$ = 5 nm (see Fig. 6a), we note the linear dependence of the velocity with the distance but we note also a small and local drop in the profile (indicated by a vertical line at the different times in the figure) followed by a small plateau. We explain this velocity drop by the fact that there where the fragmentation occurs the sheet spends a small part of the kinetic energy (therefore of the velocity) it has stored during its elongation in the creation of new fracture surfaces. Behind the fragmentation zone, we note that the spike velocity remains constant but the system stabilizes on a few nanometers. The stabilization consists in a local increase of the expansion rate because the atoms gather in order to minimize the surface energy of the new fracture created (restoring force induced by the effects of surface tension). The same phenomenon exists also with the nonlinear regimes ($h$ = 5 and 10 nm) but it is less easy to detect due to the larger range of velocity between tip and bulk (see for instance Fig. 6a).

From Fig. 10-11-12a, we can deduce that in front of the fragmentation front (between the front and the bulk), the ejected matter is still in a continuous state and described by a hydrodynamic behavior. There the velocity-distance distribution (Fig. 12a) gives information on the expansion rate that applies on the sheet. Behind this front, the matter fragments in ejecta particles, and the notion of expansion rate is no more valid. The same distribution gives now information on the particle's velocity. The mechanisms of fragmentation of the sheet were studied in detail in ref. [26]; they will be reviewed briefly in the next section.

As the fragmentation front comes close to the free surface and the bulk (around 700 ps) it stops propagating, and the effects of surface tension take place there too. In Fig. 12b, we note that for times from 700 ps to 1500 ps the local velocity drop that has propagated along the sheet stands now at the junction with the bulk (or the bottom of the bubble, this is the same). As a consequence, between the bulk (around -20 nm) and up to a distance of ~ 40 nm from the free surface, we note the existence of a region (indicated in grey tint) in which a part of the ejected matter has a velocity below $V_{FS} - |V_{Bu}|$. Since this part is slower than the bottom of the bubble (i.e. the bulk), it tends to be reabsorbed by the bulk. This is confirmed in Fig. 13 and 14 where we show a series of snapshots of the system up to a distance of 100 nm, for times between 700 ps and 1500 ps. Once the complete fragmentation of the sheet has been achieved, we observe that the particles beyond (roughly) 40 nm from the free surface tend to move away from the free surface on late times. Conversely, below this distance the particles tend to come closer to the free surface and the bulk as the time increases, and some of them begin even to be reabsorbed. This is particularly observable on the series of snapshots of Fig. 14.

In order to avoid a prohibitive time of computation, we stopped our simulations at 1500 ps. Although at this time the process of reabsorption of the slower particles is not fully fulfilled, it appears clearly that between the time 700 ps in Fig. 13 and the last time (1500 ps) in Fig. 14 the surface of the bulk has become (almost) completely flat. As for the particles, this change of morphology results from the effects of surface tension which aim at minimizing the total energy of the system. Since for the bulk we have periodic conditions in the $O_y$ and $O_z$ directions, its only configuration of minimum energy is the plane surface (for the ejected particles, the periodic conditions have no influence: therefore their shape of minimum energy is the sphere).

This evolution of morphology of the bulk allows to understand the progressive change of volume density profile observed in Fig. 7a. Indeed, as long as the volume density distribution does not exhibit a stiff profile (before typically 1300ps in Fig. 7a), it means that the sheets have not separated completely from the bulk and that the ejection process is incomplete. In other words, the system had not yet the time to minimize its total energy by making the bulk surface perfectly flat. As a consequence, the measure of the total (maximum) amount of ejected areal density does not vary anymore only on late times, as observed in Fig. 7b-7c and Fig. 8. In fact, this measure is almost fixed since the process of reabsorption of the slower particles is still in process at 1500 ps. Then we could include in the total number of ejecta some particles which will be finally reabsorbed by the bulk, but this concerns very few particles and therefore a very weak amount of areal density.

From these observation, we may also discuss the fact that the term bubble loses progressively its sense as the time increases. Indeed, it is strongly correlated to a hydrodynamic description of the ejection process (expanding RMIs). As long as the fragmentation front has not yet reached the bulk, the definition is relevant. However, when the fragmentation front reaches the bulk, the sheet separates progressively from the bulk under the effects of surface tension and the matter behavior can no more be described from a continuous point of view, using hydrodynamics. From that time, we should rather talk of a bulk distance (from the free surface) rather than a bubble amplitude, this bulk distance increasing linearly with time on late times (see Fig. 3a). For this reason, the definition of a finite and non varying amount of areal density with time is not in contradiction with an increasing boundary of integration $-|h_{Bu}(t)|$ in eq. (2). Indeed, since on late times the ejected matter has fragmented in particles which are completely separated from the bulk, even if the distance $h_{Bu}(t)$ increases, no additional mass is summed in the integral.

### B. FRAGMENTATION ZONE PROPAGATION (FZP) MODEL

From the analysis above, we may propose a model of fragmentation of the sheets; the concepts are described in Fig. 15. Let us suppose that at time $t_0$, due to a high initial expansion rate, the sheet has got thin down almost instantaneously to the critical thickness $e_c$

~ 4 nm on a given region of width $L_F$ (as observed in Fig. 9) from the point $x_i(t_0)$ to the tip of the sheet (of coordinate $h_{Sp}(t_0)$). We will address this region separately. Adjacent to this, let us consider a zone (indicated in red in Fig. 15) of width $\Delta x_{ij}(t_0) = \Delta x_0$ between the positions $x_i(t_0)$ and $x_j(t_0)$. This zone is taken as the last portion of the sheet for which the thickness $e$ is less than or equal to $e_c$ ($e > e_c$ from the free surface to $x_j(t_0)$). By neglecting the thickness variation of the jet in this zone, we may consider that the amount $\rho_0$ of matter per unit volume inside the zone is:

$$\rho_0 = \frac{m_0}{e_c \cdot \Delta x_0 \cdot L_y}, \tag{3}$$

where $L_y$ is an arbitrary length parallel to $O_y$ axis and $m_0$ is the total mass contained in the volume of the zone. In this model, we admit that once the critical thickness is reached the fragmentation and therefore the separation of the zone from the rest of the sheet is instantaneous, i.e. we neglect the characteristic time of creation, growing and percolation of the pores (a few tens of ps as reported above and in Ref. [26]).

From Fig. 6a, we can write that at a given point $x$ inside the jet and time $t$, the velocity $V(x,t)$ in the laboratory frame is:

$$V(x,t) = \dot{\varepsilon}(t) \cdot x + V_{FS}, \tag{4}$$

where $\dot{\varepsilon}(t) = V_{Sp}/h_{Sp}(t)$ is the expansion rate of the sheet previously defined and $V_{FS}$ is the free surface velocity (remember that $V_{Sp}$ is the velocity relative to $V_{FS}$ and that the spike velocity in the laboratory frame is $V_{Sp}+V_{FS}$). Therefore, at time $t_0$ the difference of velocity between the ends of the fragmentation zone is: $\Delta V_{ij}(t_0) = V(x_i(t_0)) - V(x_j(t_0)) = \dot{\varepsilon}(t_0) \cdot \Delta x_0$. From time $t_0$, the zone separates from the rest of the jet and expands self-similarly. At time $t > t_0$, its width is: $\Delta x_{ij}(t) = \Delta x_0 + \Delta V_{ij}(t_0) \cdot (t - t_0)$. As a consequence, the volume density $\rho_{ij}$ in the expanded zone at time $t$ is:

$$\rho_{ij}(t) = \frac{m_0}{e_c \cdot \Delta x_{ij}(t) \cdot L_y} = \rho_0 \cdot \frac{\Delta x_0}{\Delta x_{ij}(t)}. \tag{5}$$

While this first portion of the sheet expands self-similarly, the sheet continues to thin down which amounts to saying that the fragmentation zone counter-propagates along the sheet towards the free surface. We assume that its propagation velocity is constant. Thus, at a time $t_1 > t_0$, the fragmentation zone affects another portion of the sheet between the points $x_j(t_1)$ (= $x_j(t_0)$) and $x_k(t_1)$. We take the width of this zone constant and also equal to $\Delta x_0$: $x_j(t_1) - x_k(t_1) = \Delta x_0$. Between these two new ends, the difference of velocity is now: $\Delta V_{jk}(t_1) = V(x_j(t_1)) - V(x_k(t_1)) = \dot{\varepsilon}(t_1) \cdot \Delta x_0$. As $\varepsilon$ decreases as $1/t$, we have $\dot{\varepsilon}(t_1) < \dot{\varepsilon}(t_0)$. Therefore, when this second portion separates from the sheet, it expands self-similarly less than the first zone, and at time $t$, its volume density $\rho_{jk}$ is:

$$\rho_{jk}(t) = \frac{m_0}{e_c \cdot \Delta x_{jk}(t) \cdot L_y} = \rho_0 \cdot \frac{\Delta x_0}{\Delta x_{jk}(t)}, \tag{6}$$

with $\rho_{jk}(t) > \rho_{ij}(t) > \rho_0$. This process continues during the propagation of the zone along the sheet. By taking $\Delta x_0$ sufficiently small, the volume density profile $\rho$ at a given time $t_n$ is obtained by an almost continuous summation of the contributions $\rho_{ij} + \rho_{jk} + \ldots$ that result from the presence of the fragmentation zone along the sheet at times $t_0, t_1, \ldots$. The cumulative areal density profile is obtained by numerical integration of the volume density profile.

Our model is driven by 3 parameters: the time at which the fragmentation zone begins to propagate, its starting position and its velocity of propagation. It is important to emphasize that it applies only behind the fragmentation front.

The evolution of the region of width $L_F$ may be analyzed more simply. Since we are at the tip of the sheet, we may reasonably assume that the amount of matter contained in this region is weak and homogeneously distributed (the fragmentation has just begun). Therefore, during the self-similar expansion of this region, the resulting volume density will decrease homogeneously and it will be considered as constant (and very small) whatever the position.

In Fig. 16, we report for each roughness amplitude the profile of cumulative areal density (black curves) in linear and semi-logarithmic scales (Fig. 16a and 16b respectively) measured at the last time of our simulations (1500 ps, 1300 ps and 1100 ps for $h$ = 2, 5 and 10 nm respectively) as a function of the velocity in the cloud normalized to the free surface velocity. We report also the areal density profiles obtained with our FZP model (red curves). The parameters of the model for each roughness amplitude are reported in Table II. We note a good agreement between theoretical and simulation curves. For the linear RMI regime, the starting point of propagation of the fragmentation zone being very close to the jet tip (i.e. $L_F \sim 0$), the theoretical profile can be compared to the totality of the simulation profile. For the nonlinear regimes, the starting point is located between the free surface and the tip (i.e. $L_F \neq 0$); therefore the end of the simulation profiles cannot be fitted with the model. It is fitted by a linear function of the generic form $\alpha V/V_{FS} + \gamma$ (blue curves) since we have seen that in the region of initial width $L_F$ the volume density may be considered as constant (and the areal density is the integral of a constant).

In experiments, the areal density versus velocity distribution is generally empirically fitted by an exponential function [14]:

$$\sigma_{ejecta}(t) = \sigma_0 \exp^{-\beta(V/V_{FS}-1)} , \qquad (7)$$

where $\sigma_0$ is an empirical areal density. In Fig. 16, we compare also our profiles with such an exponential function (green curves). We note that in our linear regime of RMI ($h$ = 2 nm), the fit is also very good (see for instance the insert of fig. 16a); with the exponent $\beta \sim 14$. Therefore, the profile obtained with our FZP model in linear regime is like an exponential profile. Conversely, for the nonlinear regimes of RMI ($h$ = 5 and 10 nm), above the typical $V/V_{FS}$ ratio of 1.15, the exponential fit is not able to reproduce the simulation profile.

We may note also that if the exponential fit seems a little bit better than our FZP model for $V/V_{FS}$ < 1.15, this is mainly because it is has been specifically optimized for a given profile, obtained at a given time; but as the areal density profiles change with time, the coefficient $\beta$ changes too. Conversely, the parameters used in our model are fixed once and for all, and they do not need to be re-optimized as the time changes.

## V. EJECTA SIZE-VELOCITY DISTRIBUTION

The model we have developed is based on the self-similar expansion of different portions of the jet at different times, with less and less favourable conditions for their expansion. Let us see now the impact of this process on the volume (or size) distribution of the particles resulting from the fragmentation of these portions. In Ref [26], we studied in detail the process of fragmentation of the atom sheets. In particular, we showed that the fragmentation is driven by two basic and distinct mechanisms that occur sequentially in time. First the sheets undergo a two-dimensional (2D) mechanism of fragmentation caused by the nucleation, the growing and the coalescence of the pores of void we described earlier; and once the pores have percolated, a 2D network of ligaments of liquid metal appears. The largest of these ligaments are rather cylindrical in shape and, depending on their internal energy, they may still fragment later following a secondary 1D fragmentation mechanism that obeys a Poisson statistics. At the end of the process, all the particles become spherical and, associated to these two mechanisms, their volume distribution (number $N$ of particles with a given volume $V$) obeys on late times the generic form [26]:

$$N(V) = \eta_1 V^{-\alpha} + \eta_2(t) e^{-V/V_0}, \qquad (8)$$

with $\alpha$ = 1.15 ± 0.08; $\eta_1$ and $\eta_2$ are weighting factors. The first term, a power law, is the signature of the 2D fragmentation. Its particular property is to be scale free, i.e. it is independent of the scale at which it is observed. The second term, exponential, results from the 1D fragmentation of the largest ligaments. Unlike the power law, it is governed by a characteristic/average volume $V_0$ that depends on the scale at which it is measured. We showed that, based in particular on the model of Grady et al. [39] $V_0$ results from the competition between the internal energy of the ligaments (mainly the kinetic/strain energy stored during the expansion of the sheet) and the surface energy $\sigma \cdot A$ required to create a new fracture of surface $A$ ($\sigma$ is the surface tension of the metal).

At a distance $x$ from the free surface and time $t$, the kinetic energy of the atoms in the sheet being proportional to $\rho[\dot{\varepsilon}(t)x(t)]^2$, an estimation of $V_0$ is:

$$V_0 \propto \frac{\sigma \cdot A}{\rho[\dot{\varepsilon}(t)x(t)]^2}. \tag{9}$$

Therefore, from our study, since $\dot{\varepsilon}(t)$ decreases as $1/t$, it implies that the closer to the free surface the ligaments of liquid metal are created, the less energy they have to fragment following a secondary 1D mechanism, and the larger the average volume $V_0$ of the final particles is.

In order to illustrate this point, we concentrate our study in the following on the case $h$ = 5 nm because the effects are more obvious with a nonlinear regime due to a more important production of particles (but they exist also in the linear regime). In Fig. 17a-18a-19a, we report oblique views of the system at 200 ps, 400 ps, and 1300 ps respectively (partial view of the cloud at 1300 ps up to 700 nm), and in Fig 17b-18b-19b, we report the spatial distribution of the aggregates as a function of their volume (each point in the graphs represents one given aggregate with one given volume). As expected, we observe clearly that, as the time increases, the ejecta are progressively created from the jet tip to the free surface with an average volume (or size) which increases.

For reasons of clarity, we focus for the moment on the particles size distribution and we will describe Fig. 17c-18c-19c later.

With our binning technique, we measure the log-log volume distribution of the particles at the jet tip for the times 200 ps, 400 ps and 1300 ps in Fig. 20a and in the vicinity of the free surface (between the bulk and up to ~ 200 nm), at the last time (1300 ps) in Fig. 20b. In Fig. 20b we also report the distribution at the jet tip at 1300 ps for comparison. Then we quantify the average volume $V_0$ at each end of the cloud; and we indicate the value of $V_0$ with black dashed lines in Fig. 17b-18b-19b. We note that the ratio between the average volumes in these two extreme regions is ~ 6.6. By considering the history of propagation of the fragmentation zone in the case $h$ = 5 nm, we estimate that the particles have been created between (roughly) 200 ps and 500 ps (see for instance Fig.17a and 18a). Between these two times, the expansion rate has been divided by ~ 2.7 (from ~ 6.2×10$^9$ s$^{-1}$ to ~ 2.24×10$^9$ s$^{-1}$; see Fig. 6b), which means from eq. (9) that the theoretical average volume has been divided by a factor of ~ 7.3. This value is close to the ratio of 6.6 obtained in our simulations. In Fig. 20c, we report the total distribution of all the particles created from the bulk to the jet tip. In this case, the power law is well defined in the small volume limit because the statistics of particles is better, and in the large volume limit, the tail of the distribution is the envelope of the exponential distributions corresponding to different portions of the cloud.

In Fig 17c-18c-19c, we report the volume and areal densities of the system (in linear and semi-logarithmic scales respectively) with the theoretical curves obtained from our FZP model. We note that the volume density corresponding to the particles created at the jet tip is relatively constant and leads to a linear dependence of the areal density profile with the distance $x$ of the generic form $\alpha x + \gamma$ (magenta curves of Fig. 17c-18c-19c). This validates our hypothesis of a small and constant volume density when the thinning down of the sheet on a width $L_F$ is almost instantaneous in nonlinear regime. We note also that our model fits better the volume and areal density profiles at the last time of the computation (1300 ps) since, as emphasized in section IV.B., it applies only behind the fragmentation front (and at 1300 ps, the front has reached the free surface and the bulk).

## VI. CONCLUSION

In this paper, we investigated the mass (volume and areal densities) versus velocity as well as the size versus velocity distributions of a shock-induced cloud of particles using large scale MD simulation. A generic 3D tin crystal containing about 60×10$^6$ atoms with a sinusoidal free surface roughness was set in contact with vacuum and shock-loaded so that it melts directly on shock. Our study demonstrates the capability of MD simulations to reproduce the continuum behavior of the phenomenon, within the framework of Richtmyer-Meshkov instabilities. In particular, the bubble and spike velocities as well as the profiles and amounts of ejected volume densities are scalable at nanometric level.

Taking advantage of the possibility for MD computations to simulate fragmentation phenomena, our study shows that the atom sheets which result from the reflection of the shockwave onto the free surface progressively sharpen; and when their thickness reduces down to a critical value of a few nanometers, they become highly unstable and begin to fragment. In consequence, the distributions may be described by an analytical model based on the propagation from the tip of the RMI to the free surface of a fragmentation zone with less and less favourable conditions of self-similar expansion and fragmentation.

These processes are mainly driven by the time, the expansion rate of the sheet and the surface tension of the metal. A lot of work still remains to do to transpose the results to the macroscopic/experimental world, and in particular, the scaling in time needs to be thoroughly studied. It will be addressed in a future work. Nevertheless, the associated analysis may bring some line of thoughts for future experiments or models.

In particular, our simulations show that (i) the particles are not created at the same time (the small and fast particles are created in the early times at the jet tip, and the large and slow ones are created in the late times near the free surface) and (ii) at a given time, the strain rate is constant over the sheet in its longitudinal direction. This is in contradiction with the common postulates that (i) the particles are created instantaneously at the reflection of the shockwave onto the free surface [11, 15] and (ii) the strain rate is spatially distributed over the depth of the jet [16, 20]. Moreover, our simulations show that even with a single perturbation wavelength, the particles are distributed in sizes, unlike the assumption in Ref. [12] that the distribution of the particles in sizes is due to the distribution in wavelengths (modes) of the initial surface perturbation (each mode/wavelength producing a typical size of particle). All these discrepancies are the consequence of the modeling of the phenomenon with a fragmentation zone that propagates.

The simulations show also that the surface tension effects need to be accurately characterized and modeled since they drive not only the size of the particles but also the mechanism of separation of the ejected sheet from the bulk (minimization of the energy). This mechanism influences the temporal evolution of the volume and areal density profiles and therefore the determination of the total amount of areal density ejected.

The typical particle average volume $V_0$ that results from our computations is about $10^6$ Å$^3$. From eq. (9), if we consider that the expansion rate involved in experiments is between 3 or 4 orders of magnitude smaller than ours, then the average volume of the particles at experimental scale should be larger than our MD volume by a factor of $10^6$-$10^8$, i.e. $V_0$ should be of the order of $10^{12}$-$10^{14}$ Å$^3$. This corresponds to particle radii of about 1-10 µm which are compatible with the first size measurements [21-23]. The determination of an average particle volume/size based on an energy balance principle seems therefore relevant.

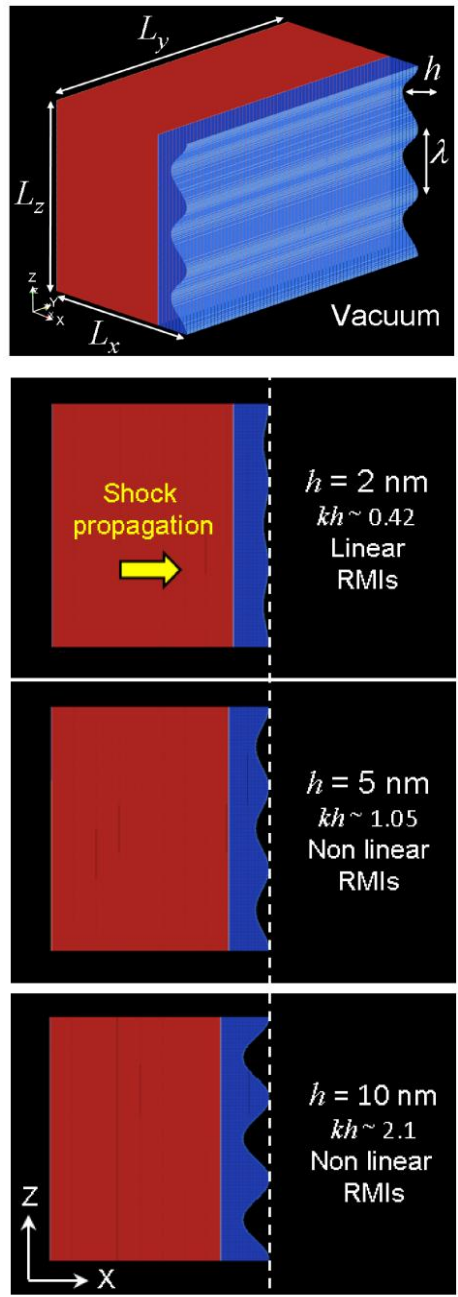

Figure 1: Description of the generic system simulated. The white dashed line indicates the position of the free surface at the beginning of the simulations.

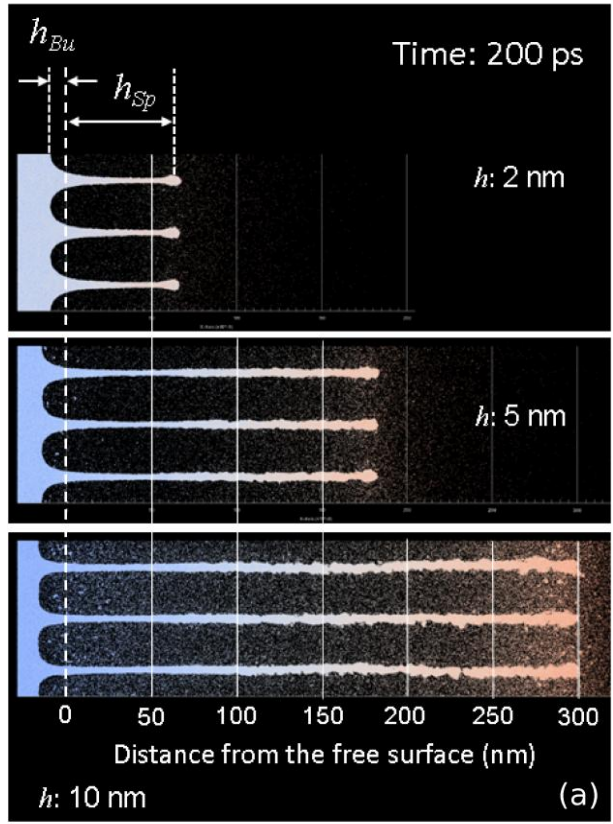

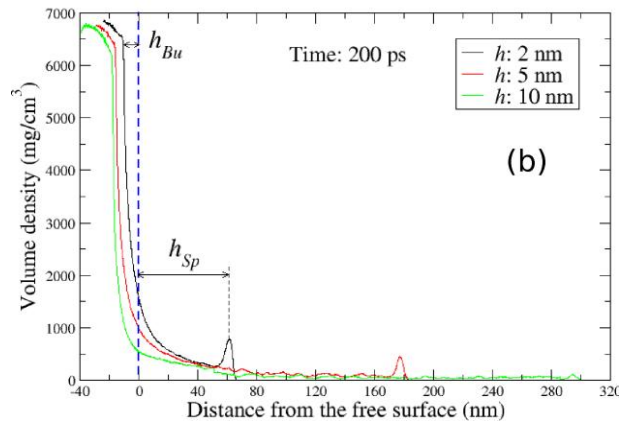

Figure 2: (a) Side views of the ejected sheets for the 3 roughness amplitudes at 200 ps and (b) Corresponding profile of volume density as a function of the distance from the free surface.

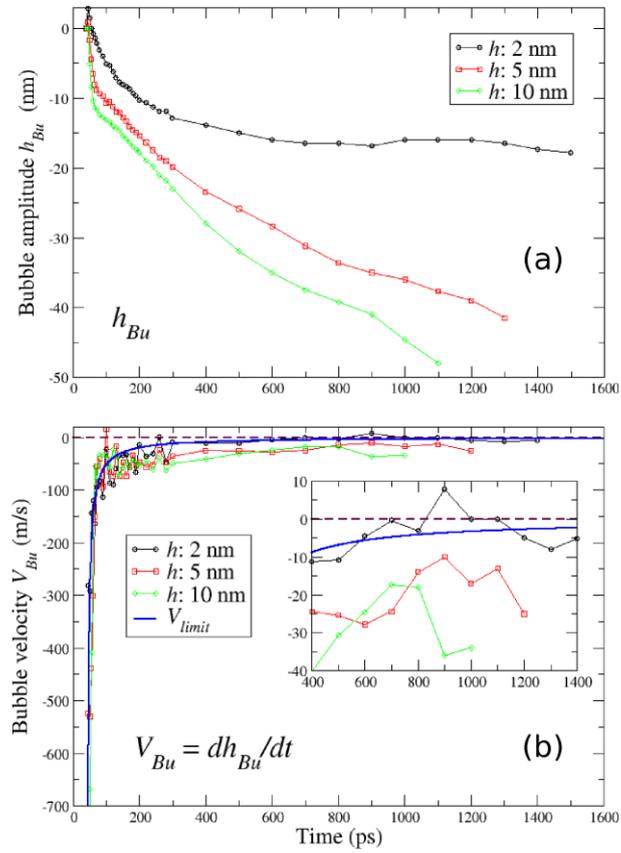

Figure 3: Temporal evolution of (a) the bubble amplitude and (b) the bubble velocity for the 3 roughness amplitudes.

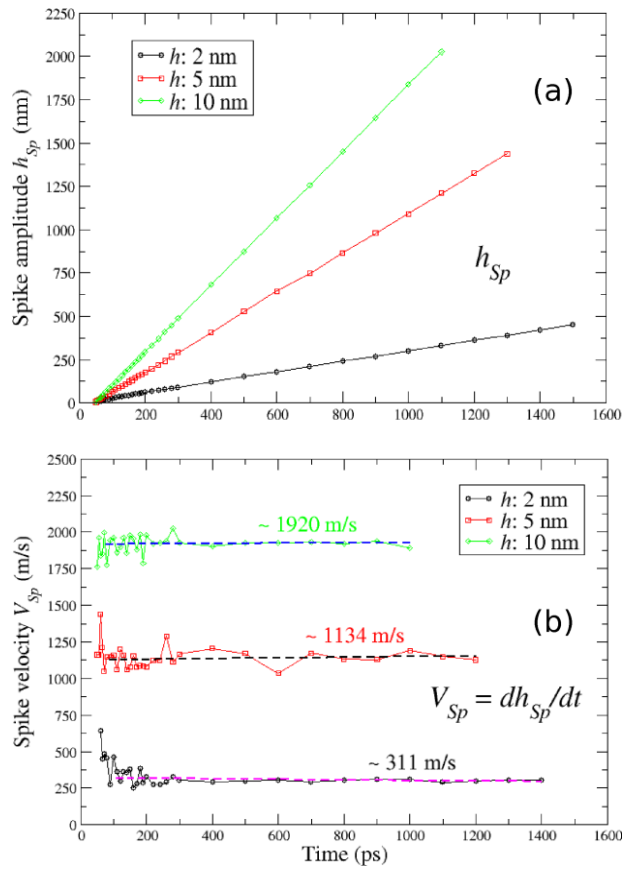

Figure 4: Temporal evolution of (a) the spike amplitude and (b) the spike velocity for the 3 roughness amplitudes.

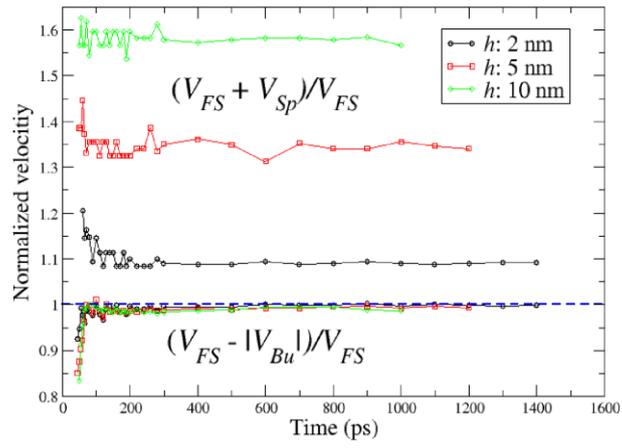

Figure 5: Temporal evolution of the bubble and spike velocities in the laboratory frame normalized to the free surface velocity for the 3 roughness amplitudes.

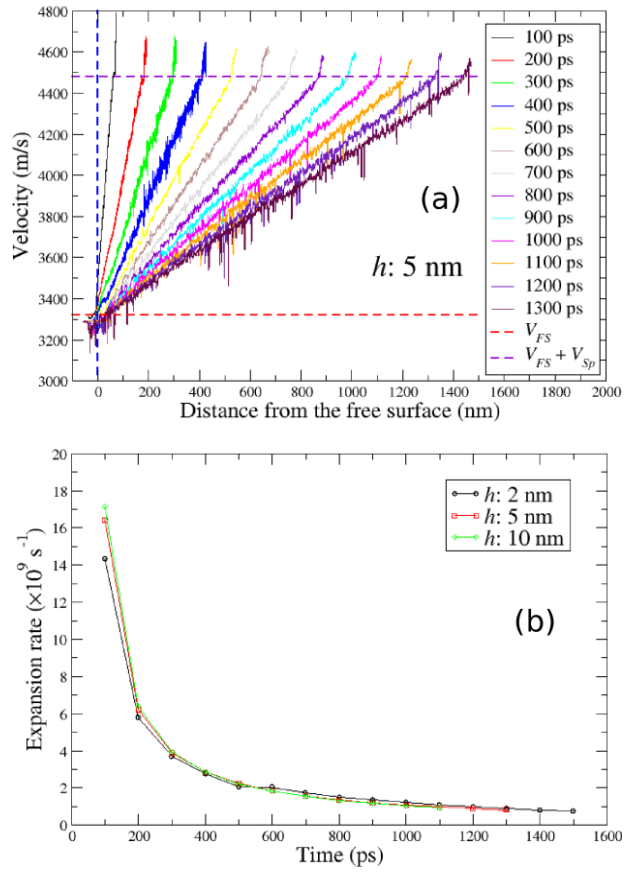

Figure 6: (a) Velocity profile inside the sheet as a function of the distance from the free surface for the roughness amplitude h = 5 nm, and (b) Temporal evolution of the expansion rate for the 3 roughness amplitudes.

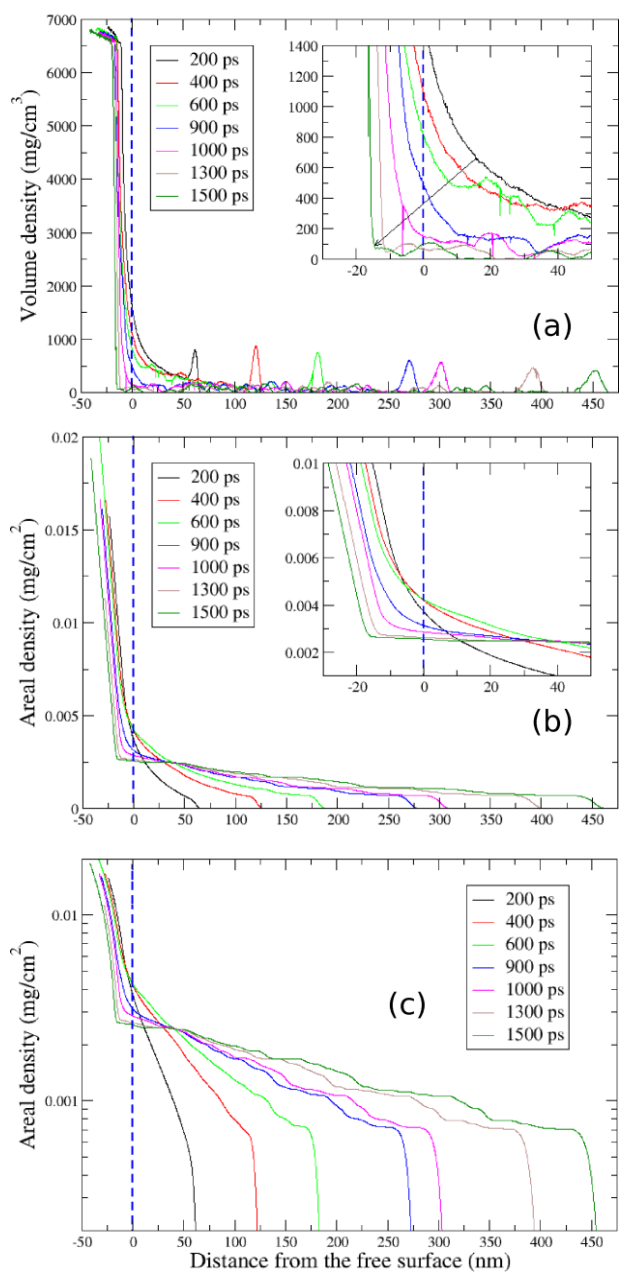

Figure 7: Profiles of (a) volume density, (b) areal density in linear scale and (c) areal density in semi-logarithmic scale as a function of the distance from the free surface for the roughness amplitude $h = 2$ nm.

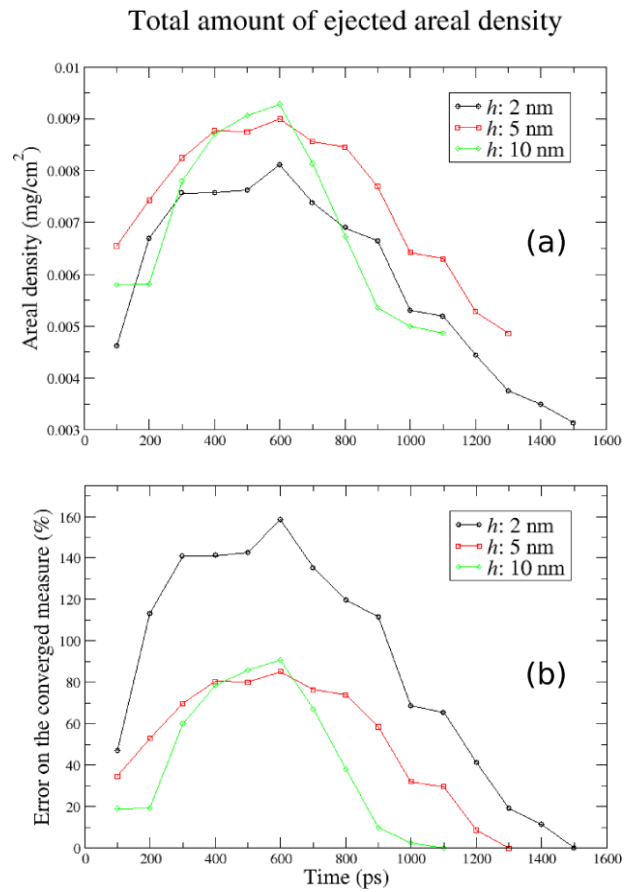

Figure 8: (a) Temporal evolution of the total amount of ejected areal density measured at the bottom of the jet and (b) error of the measure on the value obtained at the last time, for the 3 roughness amplitudes.

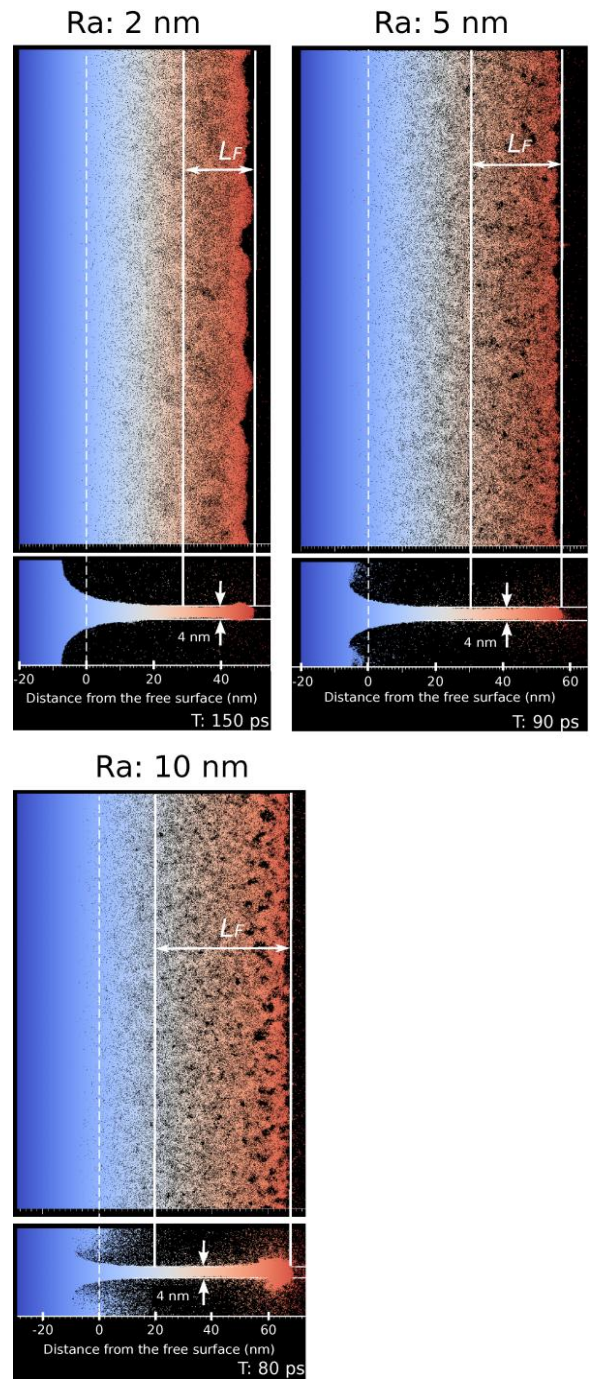

Figure 9: Top and side views of the upper ejected sheet for the 3 roughness amplitudes at the moment where the sheet begins to become unstable.

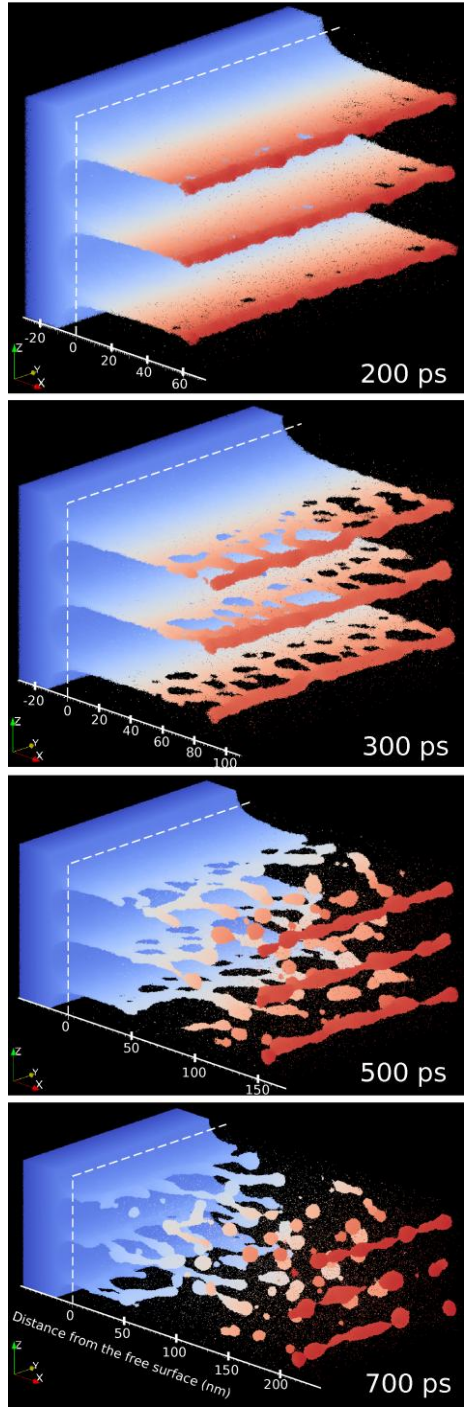

Figure 10: Oblique views of the system for the roughness amplitude h = 2 nm between times 200 ps and 700 ps.

Roughness amplitude: 2 nm

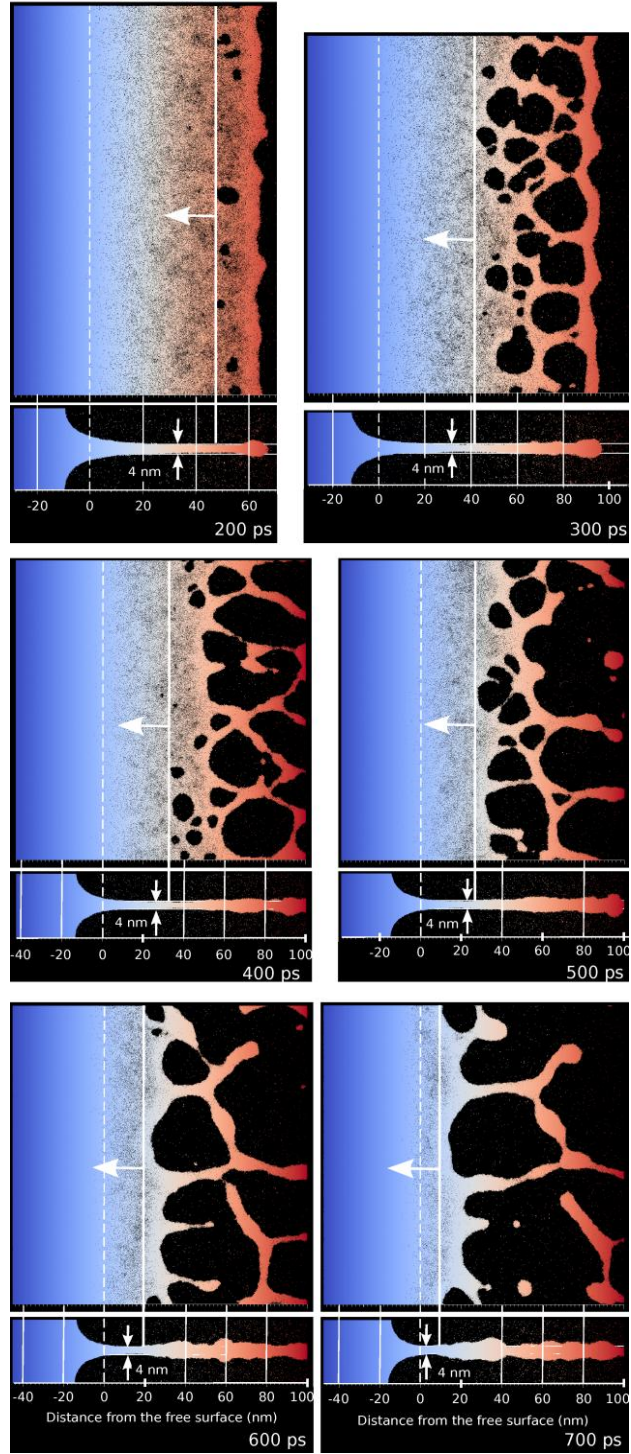

Figure 11: Top and side views of the upper sheet for the roughness amplitude h = 2 nm between times 200 ps and 700 ps.

Figure 12: Velocity profile inside the sheet as a function of the distance from the free surface for the roughness amplitude h = 2 nm (a) for times between 200 ps and 700 ps and (b) for times between 700 ps and 1500 ps.

Roughness amplitude: 2 nm

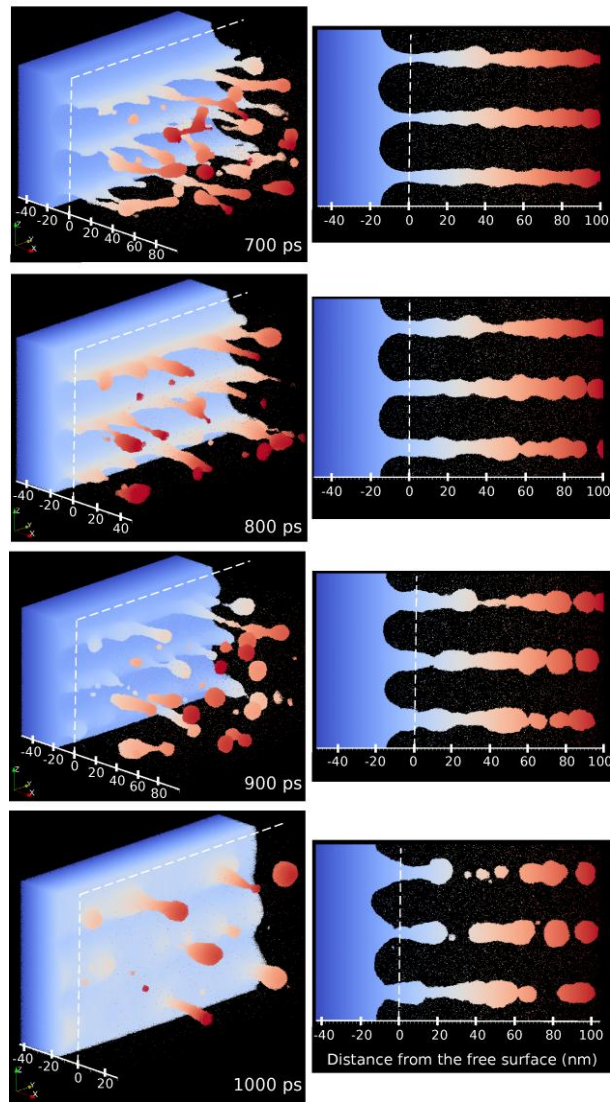

Figure 13: Oblique and side views of the system near the free surface for the roughness amplitude h = 2 nm and for times between 700 ps and 1000 ps.

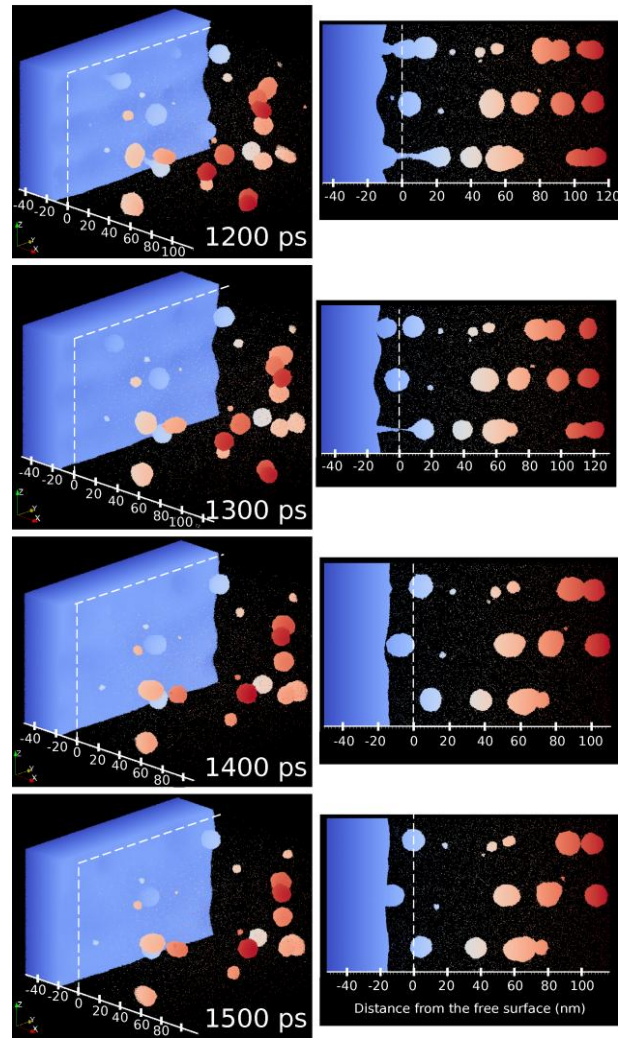

Figure 14: Oblique and side views of the system near the free surface for the roughness amplitude h = 2 nm and for times between 1200 ps and 1500 ps.

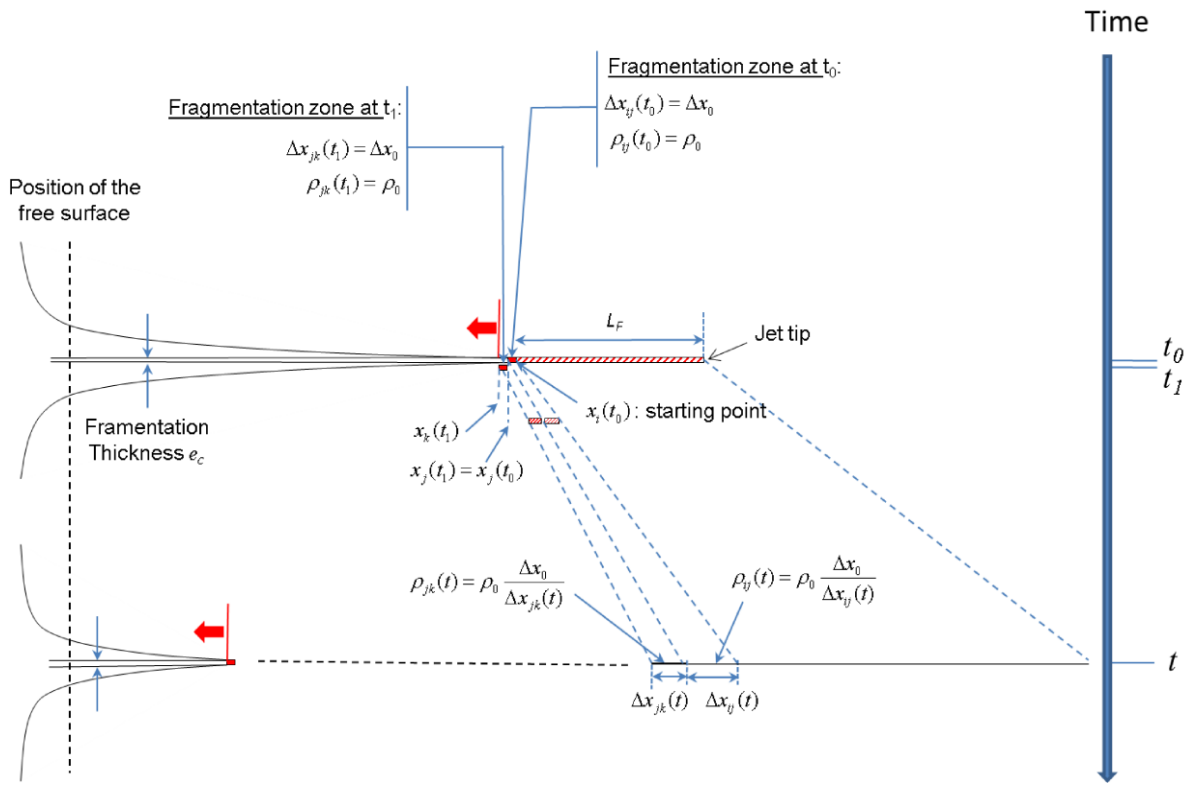

Figure 15: Principle of the fragmentation zone propagation (FZP) model.

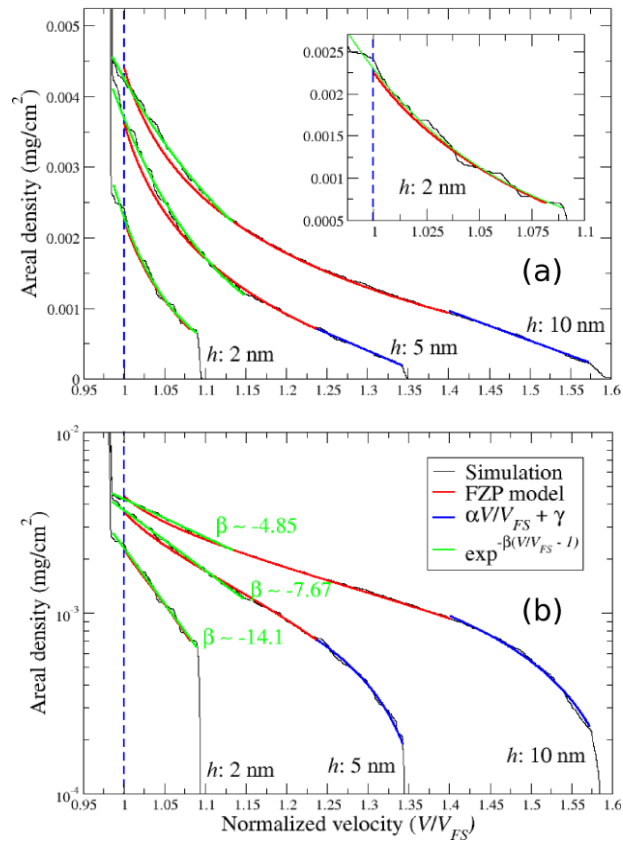

Figure 16: Comparison of the areal density profiles (a) in linear and (b) in semi-logarithmic scales measured at the last time of computation for the 3 roughness amplitudes with the profiles obtained with the FZP model and exponential function.

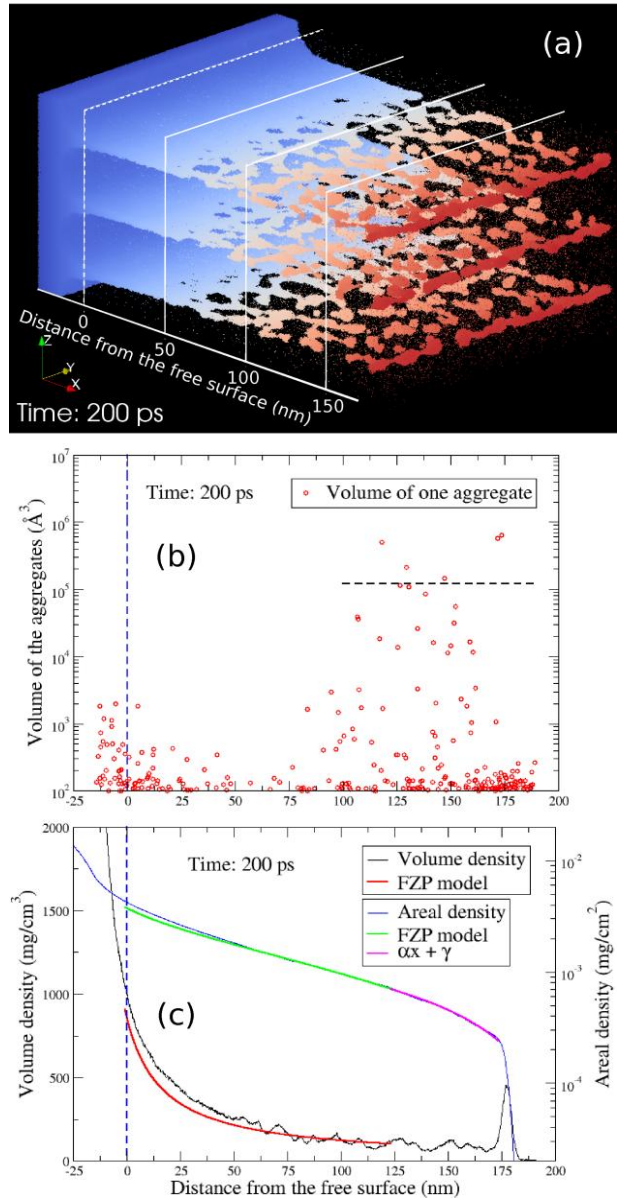

Figure 17: For the roughness amplitude h = 5 nm at time 200 ps (a) oblique view of the system, (b) spatial distribution of the aggregates following their volume and (c) volume and areal density distributions (in linear and semi-logarithmic scales respectively) obtained with the simulations and the FZP model as a function of the distance from the free surface.

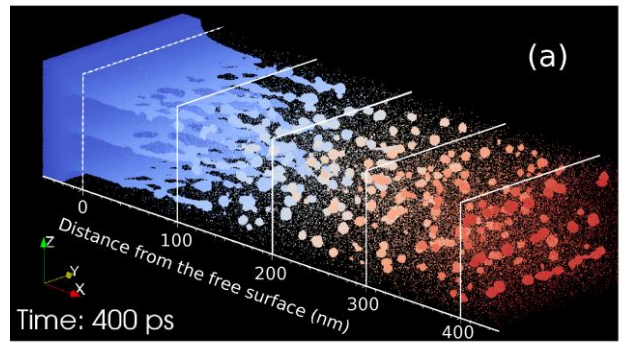
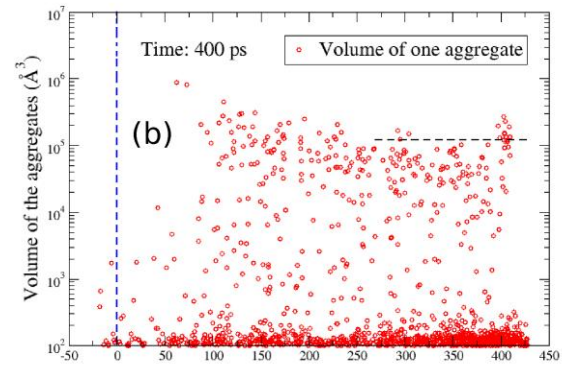
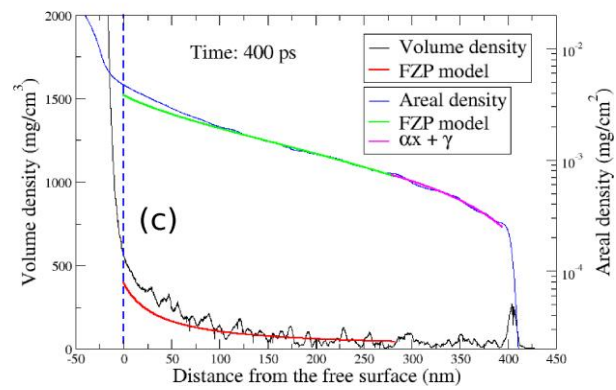

Figure 18: For the roughness amplitude h = 5 nm at time 400 ps (a) oblique view of the system, (b) spatial distribution of the aggregates following their volume and (c) volume and areal density distributions (in linear and semi-logarithmic scales respectively) obtained with the simulations and the FZP model as a function of the distance from the free surface.

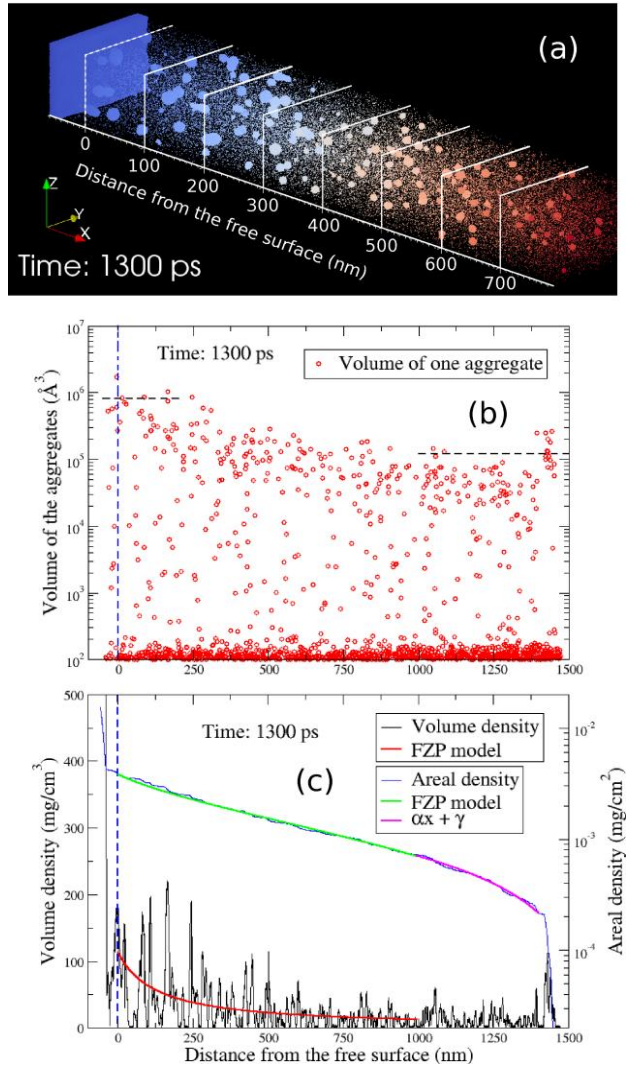

Figure 19: For the roughness amplitude h = 5 nm at time 1300 ps (a) oblique view of the system, (b) spatial distribution of the aggregates following their volume and (c) volume and areal density distributions (in linear and semi-logarithmic scales respectively) obtained with the simulations and the FZP model as a function of the distance from the free surface.

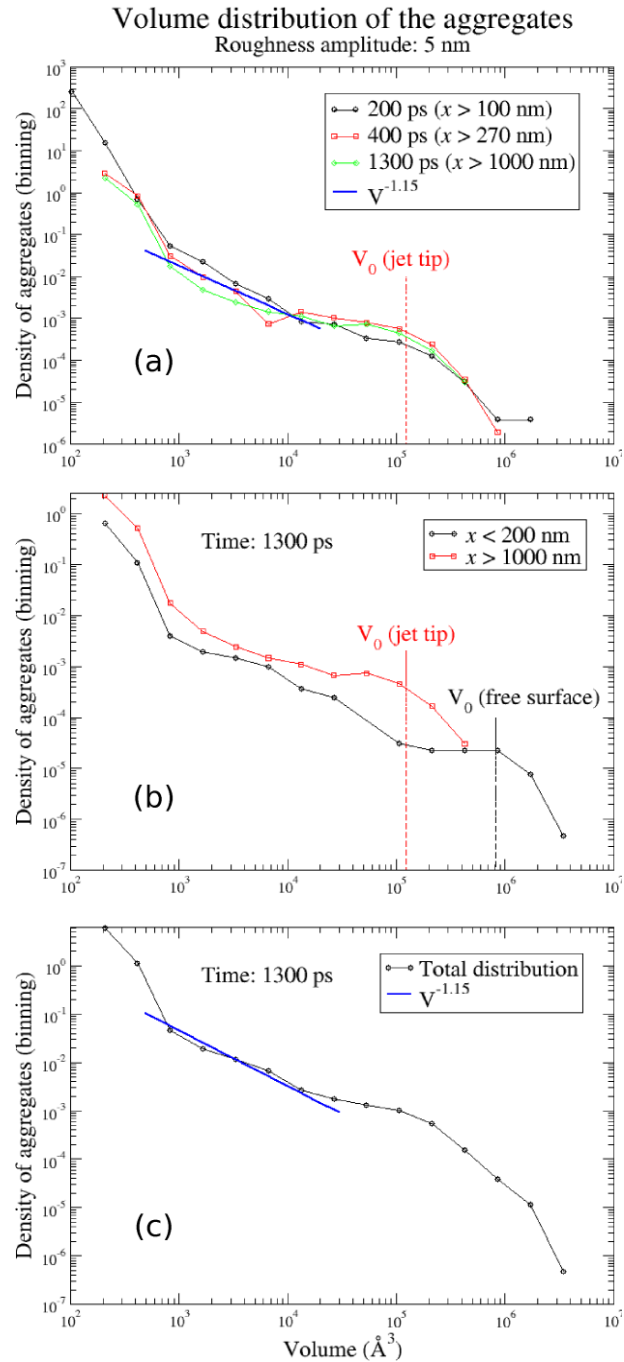

Figure 20: Log-log volume distribution of the particles for the roughness amplitude h = 5 nm (a) at the jet tip between times 200 ps and 1300 ps, (b) at the free surface and the jet tip at 1300 ps and (c) in the whole cloud at 1300 ps.

**TABLES**

**Table I**

|  | Tin (Sn) |
|---|---|
| Compression ratio: $V_1/V_0$ | 0.70 |
| Velocity particle: $u_p$ (m/s) | ~ 1570 |
| Pressure: $P_1$ (GPa) | ~ 65 |
| Temperature: $T_1$ (K) | ~ 2100 |

Thermodynamic properties of the shock-loaded Sn crystal

**Table II**

| Roughness amplitude | $h$ : 2 nm | $h$ : 5 nm | $h$ : 10 nm |
|---|---|---|---|
| Time $t_0$ of beginning of propagation of the fragmentation zone (ps) | 150 | 100 | 80 |
| $h_{Sp}$ at $t_0$ (nm) | 45 | 66 | 66 |
| Position of the fragmentation zone at $t_0$ (nm) | 38 | 45 | 45 |
| Propagation velocity of the fragmentation zone relative to $V_{FS}$ (m/s) | -80 | -100 | -110 |

Parameters of the FZP model for each roughness amplitude